# ELECTRON CLOUD TRAPPING IN RECYCLER COMBINED FUNCTION DIPOLE MAGNETS

S. A. Antipov, The University of Chicago, Chicago, IL 60637, USA
S. Nagaitsev, Fermilab, Batavia, IL 60510, USA

*Abstract*

Electron cloud can lead to a fast instability in intense proton and positron beams in circular accelerators. In the Fermilab Recycler the electron cloud is confined within its combined function magnets. We show that the field of combined function magnets traps the electron cloud, present the results of analytical estimates of trapping, and compare them to numerical simulations of electron cloud formation. The electron cloud is located at the beam center and up to 1% of the particles can be trapped by the magnetic field. Since the process of electron cloud build-up is exponential, once trapped this amount of electrons significantly increases the density of the cloud on the next revolution. In a Recycler combined function dipole this multi-turn accumulation allows the electron cloud reaching final intensities orders of magnitude greater than in a pure dipole. The multi-turn build-up can be stopped by injection of a clearing bunch of $10^{10}$ p at any position in the ring.

## FAST INSTABILITY

In 2014 a fast transverse instability was observed in the proton beam of the Fermilab Recycler. The instability acts only in the horizontal plane and typically develops in about 20-30 revolutions. It also has the unusual feature of selectively impacting the first batch above the threshold intensity of ~ $4*10^{10}$ protons per bunch (Fig. 1). These peculiar features suggest that a possible cause of the instability is electron cloud. Earlier studies [1, 2] indicated the presence of electron cloud in the ring and suggested the possibility of its trapping in Recycler combined function magnets.

The fast instability seems to be severe only during the start-up phase after a shutdown, with significant reduction being observed after beam pipe conditioning during beam scrubbing runs [3]. It does not limit the current operation with slip-stacking up 700 kW of beam power, but may pose a challenge for a future PIP-II intensity upgrade.

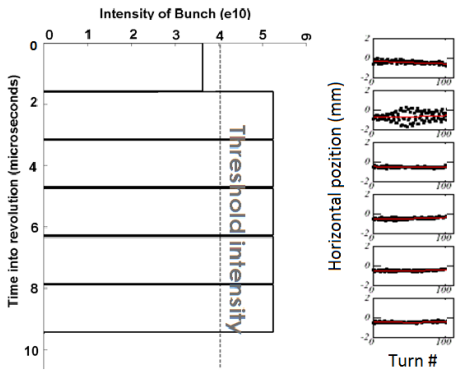

Figure 1: The first batch above the threshold intensity suffers the blow-up after injection into the ring [3].

## ELECTRON CLOUD TRAPPING

In a combined function dipole the electrons of the cloud move along the vertical field lines. This motion conserves their energy $E$ and magnetic moment

$$\mu = \frac{mv_\perp^2}{2B} = const, \quad (1)$$

where $v_\perp$ is the component of the velocity normal to the magnetic field $B$. As an electron moves closer to a magnet pole it sees a higher $B$ (Fig. 2) and it can reflect back if

$$E - \mu B = 0 \quad (2)$$

Alternatively, the electron will reflect back at the point of maximum magnetic field if the angle between the electron's velocity and the field lines is greater than:

$$\theta > \theta_{max} = \cos^{-1}(\sqrt{B_0 / B_{max}}). \quad (3)$$

Particles with angles $\theta_{max} < \theta \le \pi/2$ are trapped by magnetic field. For Recycler magnets (Table 1), Eq. (3) gives a capture of ~$10^{-2}$ particles of electron cloud, assuming uniform distribution.

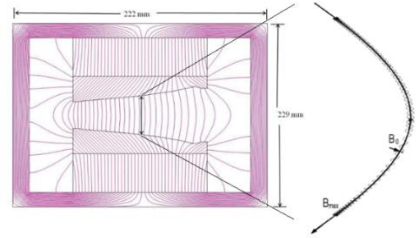

Figure 2: Electron cloud can get trapped by magnetic field of a combined function magnet.

## SYMPLIFIED ANALYTICAL MODEL

### Electron capture by the field

Let us look at the process of electron cloud trapping in more detail and consider the last two bunches of the batch. The first bunch kicks the electrons of the cloud, created by the batch. With an energy of the order of 100 eV the electrons drift along the magnetic field in the vacuum chamber, finally reaching its walls and producing secondary electrons with the energies of a few eV [4]. In the absence of the beam these secondary electrons would eventually reach the aperture and die. But the next proton bunch can stop a fraction of the secondaries, reducing their angle to $\theta < \theta_{max}$ (Fig. 3). These electrons will remain trapped in the magnetic field after the beam is gone.

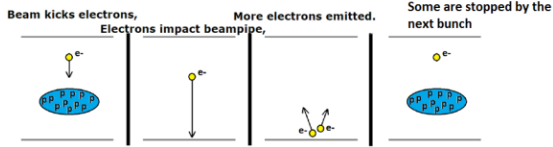

Figure 3: The first bunch kicks the cloud, creating secondary electrons, the second bunch stops some of them.

Thanks to the strong magnetic field the motion of an electron is essentially a 1D problem that can be easily solved. Initially, the secondaries start at the wall of the vacuum chamber, at $y = -A/2$, and have some distribution of velocities $\mathbf{v} = (v_y, v_\perp)$. During one RF period the electrons travel $\Delta y = v_y \tau_{RF}$ before receiving a kick from the next proton bunch. Assuming a transverse waterbag and longitudinal Gaussian beam profile and neglecting the self space-charge of the electron cloud the equation of motion becomes

$$\begin{cases} y'' + C \cdot y \exp(-t^2 c^2 / 2\sigma_s^2) = 0, & y < r_0 \\ y'' + C \cdot \dfrac{r_0^2}{y} \exp(-t^2 c^2 / 2\sigma_s^2) = 0, & y \geq r_0 \end{cases} \quad (4)$$

$$C = \frac{q_e^2 N_b}{m_e r_0^2 \sqrt{2\pi \sigma_s^2}},$$

where $r_0$ and $\sigma_s$ are the transverse and longitudinal bunch sizes and $N_b$ – bunch population.

For some initial values of $v_y$ an electron loses the vertical velocity completely after the kick by the proton. For Fermilab Recycler (Table 1) we find five such trapping 'modes' (Fig. 4).

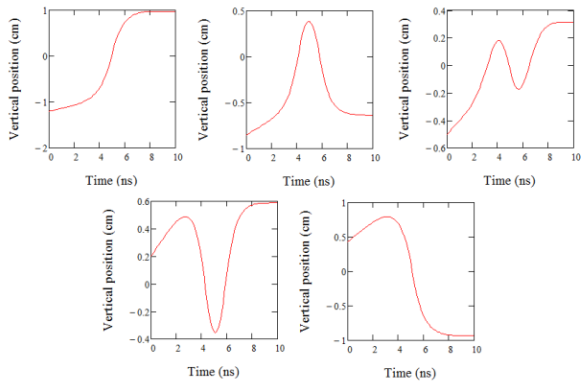

Figure 4: There are five distinct trapping 'modes' in Recycler combined function dipoles. $E = 5$ eV.

Assuming an initial distribution from [4] one can estimate the fraction of the electron cloud that stays trapped by the field after the beam is gone. For Recycler magnets this ratio can be as high as a few percent for low beam intensities. At the current operational intensity of $5 \times 10^{10}$ ppb about 1% of particles are trapped, and at higher intensities this amount gets smaller (Fig. 5).

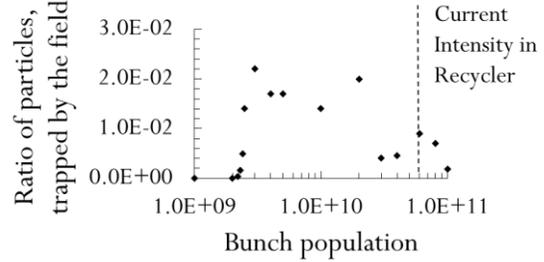

Figure 5: About 1% of electrons are trapped at the current beam intensity in Recycler. This fraction reduces for higher intensities.

*Lifetime of the trapped cloud*

Long-term confinement of the electron cloud can be affected by two effects: longitudinal drift and scattering. The drift is caused by the absence of magnetic field gradient in the longitudinal direction. The longitudinal drift velocity is

$$v_d = \frac{1}{2} \omega_c r_c^2 \frac{B'}{B_0}, \quad (5)$$

where $\omega_c$ is the cyclotron frequency and $r_c$ – the radius of the orbit. In one revolution period the cloud travels the distance $l_d = v_d T_0$, and if $l_d \sim L_{dipole}$ it may escape the magnet and decay. For the parameters of Fermilab Recycler $v_d < 2 \times 10^5$ cm/s and $l_d < 2$ cm, so the drift can be neglected.

The lifetime of trapped electrons may be also limited by scattering on each other and residual gas. The Coulomb scattering cross-section can be estimated as

$$\sigma_C = \frac{4\pi}{9} \frac{e^4 \ln \Lambda}{(kT)^2} \sim 10^{-15} \text{cm}^2 \quad (6)$$

The inelastic scattering cross-section for the energies in question is also of the order of $10^{-15}$ cm$^2$ [5]. Combining the two effects we obtain a lifetime ~ 10 ms for the electron cloud density $n_e < 10^7$ cm$^{-3}$ and the residual gas pressure $p \sim 10^{-8}$ Torr. Since the resulting lifetime of the electron cloud is much larger than the revolution period of 11 μs, all the trapped cloud will be present on the next turn.

*Electron cloud clearing with a witness bunch*

As mentioned above, the trapping requires at least two bunches: the first to kick the cloud and create the secondaries; and the second to stop a fraction of those. Therefore, a single bunch of high enough intensity does not trap the cloud but clears the aperture instead. This clearing bunch can be used to indicate the presence of the trapped electron cloud and measure its density [6] or to bring the electron cloud density below the threshold, stabilizing the beam.

## NUMERICAL SIMULATION OF ELECTRON CLOUD BUILD-UP

We simulated electron cloud build-up over multiple revolutions in a Recycler dipole using the PEI code [7]. For a pure dipole field, the cloud rapidly builds up during the passage of the bunch train and then decays back to the initial ionization electron density in about 300 RF buckets, or ~ 6 µs (Fig. 6). When the field gradient is added, up to 1% of the electron cloud stays trapped, increasing the initial density on the next revolution. The final density, which the cloud reaches after ~ 10 revolutions, is two orders of magnitude greater than in the pure dipole case (Fig. 6). The resulting cloud distribution is a stripe along the magnetic field lines, with higher particle density being closer to the walls of the vacuum chamber (Fig. 7). The width of the stripe is approximately equal to the size of the beam and its intensity increases from turn to turn as the cloud builds up.

At lower densities ~$10^{-2}$ of particles are trapped, which agrees with the analytic estimate (Fig. 5); as the density of electron cloud increases the trapping ratio goes down to ~$10^{-3}$, probably due to the space charge of electron cloud.

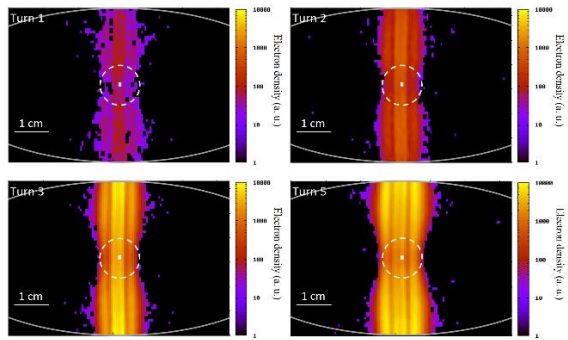

Figure 7: Electron cloud forms a stripe inside the vacuum chamber and its intensity increases with the number of turns. Its horizontal position – beam center (white dot). White circles represent 2 rms beam size.

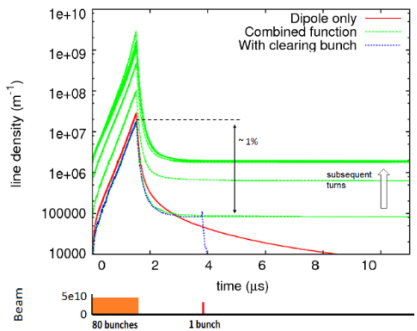

Figure 6: In a combined function magnet the electron cloud accumulates over many revolutions, reaching much higher line density, than in a dipole. A clearing bunch destroys the trapped cloud, preventing the accumulation.

A bunch of $5 \times 10^{10}$ protons, added 120 RF buckets after the main batch, destroys the trapped cloud, preventing the multi-turn build-up (Fig. 6). First, one can see a small increase in the cloud density as the clearing bunch kicks the cloud and it reaches the vacuum chamber, producing the secondary electrons. Then, the density rapidly drops as these secondaries reach the aperture.

Table 1: Recycler parameters for simulation in PEI

| | |
|---|---|
| Beam energy | 8 GeV |
| Machine circumference | 3.3 km |
| Batch structure | 80 bunches, 5e10 p |
| Tunes: x, y, z | 25.45, 24.40, 0.003 |
| RF harmonic, period | 588; 18.9 ns |
| RMS bunch size: x, y, z | 0.3, 0.3, 60 cm |
| Secondary emission yield | 2.1 @ 250 eV |
| Density of ionization e- | $10^4$ m$^{-1}$ (at $10^{-8}$ Torr) |
| B-field and its gradient | 1.38 kG, 3.4 kG/m |
| Magnet length | 5 m |
| Beampipe | Elliptical, 100 x 44 mm |

## CONCLUSION

Combined function magnets are widely used in the present day machines. Because of the gradient of the magnetic field (which provides the focusing) the electron cloud can be trapped in the magnetic field of such magnets. These trapped particles make it possible for the cloud to accumulate over multiple revolutions, possibly leading to a fast transverse instability.

We have created an analytical model that allows the estimation of the amount of the cloud captured in the magnet. We have shown that up to 1% of the electron cloud can be trapped in the magnetic field of combined function magnets of FNAL Recycler. This fraction of trapped particles will go down for higher intensities in Recycler.

Numerical simulation in PEI agrees with the analytical estimate and confirms that the trapping significantly affects the density of the electron could. It allows the cloud to accumulate over multiple revolutions reaching a density much higher than in a pure dipole. For the parameters of Fermilab Recycler with one batch of normal intensity the cloud reaches ~ $10^9$ m$^{-1}$ in a combined function magnet compared to ~ $10^7$ m$^{-1}$ in a dipole of the same field strength. An addition of a clearing bunch destroys the trapped cloud, preventing the multi-turn accumulation.

## ACKNOWLEDGMENT

The authors are grateful to K. Ohmi (KEK) for his help with PEI code. Fermilab is operated by Fermi Research Alliance, LLC under Contract No. DE-AC02-07CH11359 with the United States Department of Energy.